# Ultra-clean high-mobility graphene on technologically relevant substrates

Ayush Tyagi[ab], Vaidotas Mišeikis*[be], Leonardo Martini[b], Stiven Forti[b], Neeraj Mishra[be], Zewdu M. Gebeyehu[be], Marco A. Giambra[c], Jihene Zribi[d], Mathieu Frégnaux[d], Damien Aureau[d], Marco Romagnoli[c], Fabio Beltram[a], Camilla Coletti*[be]

**Author affiliations**

* Corresponding authors

[a] NEST, Scuola Normale Superiore, Piazza San Silvestro 12, 56127 Pisa, Italy

[b] Center for Nanotechnology Innovation @NEST, Istituto Italiano di Technologia, Piazza San Silvestro 12, 56127 Pisa, Italy

[c] Photonic Networks and Technologies Lab, CNIT, 56124 Pisa, Italy; CamGraPhIC srl, 56124 Pisa, Italy.

[d] Institut Lavoisier de Versailles UMR 8180 Université Paris-Saclay, UVSQ, CNRS, 78035 Versailles, France

[e] Graphene Labs, Istituto Italiano di Tecnologia, via Morego 30, 16163 Genova, Italy

E-mail: vaidotas.miseikis@iit.it; camilla.coletti@iit.it

**Abstract**

Graphene grown via chemical vapour deposition (CVD) on copper foil has emerged as a high-quality, scalable material, that can be easily integrated on technologically relevant platforms to develop promising applications in the fields of optoelectronics and photonics. Most of these applications require low-contaminated high-mobility graphene (i.e., approaching 10 000 cm$^2$ V$^{-1}$ s$^{-1}$ at room temperature) to reduce device losses and implement compact device design. To date, these mobility values are only obtained when suspending or encapsulating graphene. Here, we demonstrate a rapid, facile, and scalable cleaning process, that yields high-mobility graphene directly on the most common technologically relevant substrate: silicon dioxide on silicon (SiO$_2$/Si). Atomic force microscopy (AFM) and spatially-resolved X-ray photoelectron spectroscopy (XPS) demonstrate that this approach is instrumental to rapidly eliminate most of the polymeric residues which remain on graphene after transfer and fabrication and that have adverse effects on its electrical properties. Raman measurements show a significant reduction of graphene doping and strain. Transport measurements of 50 Hall bars (HBs) yield hole mobility $\mu_h$ up to ~9000 cm$^2$ V$^{-1}$ s$^{-1}$ and electron mobility $\mu_e$ up to ~8000 cm$^2$ V$^{-1}$ s$^{-1}$, with average values $\mu_h$~7500 cm$^2$ V$^{-1}$ s$^{-1}$ and $\mu_e$~6300 cm$^2$ V$^{-1}$ s$^{-1}$. The carrier mobility of ultraclean graphene reach values nearly double of that measured in graphene HBs processed with acetone cleaning, which is the method widely adopted in the field. Notably, these mobility values are obtained over large-scale and without encapsulation, thus paving the way to the adoption of graphene in optoelectronics and photonics.

## 1. Introduction

In the last years, graphene has shown its potential in numerous technological applications because of its many useful properties such as high electrical and thermal conductivity.[1,2] In particular, thanks to tremendous progress made in the field of scalable graphene synthesis *via* chemical vapour deposition (CVD), wafer-scale graphene is now accessible and ready to be integrated for different applications in fields ranging from photonics, to optoelectronics, to sensing.[3–8] Most of these applications require high-mobility ultra-clean graphene directly on a technologically-relevant substrate such as silicon dioxide on silicon ($SiO_2$/Si). In particular, photonic devices with performance that is competitive with that of conventional technology require graphene with charge-carrier mobility near 10000 $cm^2$ $V^{-1}$ $s^{-1}$ at carrier density $\sim 10^{12}$ $cm^{-2}$ [9] in order to improve Seebeck coefficient in photothermal effect detectors[10] and extinction ratio in photonic electro-absorption modulators.[9–11] High mobility is also desirable to limit propagation losses and allow for reduced geometrical footprint.[9] Also, low contamination is a requirement of foundries in which CVD graphene is included in integration process flows. The contaminant threshold for backend of line in a CMOS fab is $10^{12}$ at/$cm^2$ whereas in the frontend of line the threshold is two orders of magnitude more stringent.[12,13]

Since state-of-the-art (SOTA) scalable graphene is presently obtained via chemical vapour deposition (CVD) on metal substrates,[8] the standard fabrication of graphene devices requires: (i) an unavoidable transfer step involving coating the graphene with polymeric resists (which acts as the support layer during the transfer) and (ii) optical or e-beam lithography (EBL). Polymethyl methacrylate (PMMA), in particular, is widely used for fabrication as well as transfer[14] of CVD graphene. A well-known issue in graphene processing is the presence of PMMA residues on graphene due to strong physical and chemical adsorption effects.[15] Owing to the monolayer nature of graphene, surface adsorbents can induce carrier scattering, thus reducing the resulting mobility.[16] To realize high-performing opto-electronic and photonic devices of technological relevance, flat and contaminant-free graphene over large-scales is essential. Various methods have been used to address the issue of the polymer contamination on graphene, including stencil mask lithography,[17] mechanical cleaning with the tip of an atomic force microscope (AFM),[18,19] current-induced cleaning,[20] PMMA degradation by laser treatment,[21,22] high-temperature annealing[23–25] and wet chemical cleaning,[23,26,27] though each of these presents its own drawbacks. Stencil mask lithography relies on a physical mask placed in close proximity to the sample to define the metallic contacts or an etching pattern in graphene. While this does not require subjecting graphene to any polymer, the fragility of the masks imposes a compromise between the size of the patterning area and resolution. Furthermore, it does not allow the flexibility offered by EBL for rapid device prototyping. An effective cleaning of polymer residues from the graphene surface was demonstrated by "sweeping" it with an AFM tip operated in contact mode, however, this method is constrained to clean local areas only (typically, on the order of tens of microns) and is very time-consuming. Similar constraints apply to current- and laser-induced cleaning. Thermal annealing is compatible with large-scale processing, but, when performed on graphene/Si-$SiO_2$, it was shown to increase doping and decrease mobility by inducing strong interactions between graphene and the substrate.[24,28] To date, wet chemical cleaning is the most adopted approach to prepare graphene prior to device implementation.[23] However, the best values of electron and hole mobility obtained for CVD graphene over wafer-scale to date (i.e., not one device performance) are limited to ≤ 4800 cm$^2$ V$^{-1}$ s$^{-1}$.[8,29] Strong solvents such as *N*-methyl-2-pyrrolidone (NMP) are sometimes used for complete removal of polymer residues from graphene, and they can induce lattice defects.[30]

In this work, we demonstrate that by using a two-step wet chemical process, graphene cleanliness and electrical performance is significantly improved with respect to other chemical treatments used so far.[14,23] We perform a systematic comparison of CVD-grown graphene processed with standard single-step cleaning (1SC) in acetone and two-step cleaning (2SC) in acetone and remover AR 600-71, the latter being a two-component solvent. We use atomic force microscopy (AFM), Raman spectroscopy, X-ray photoelectron spectroscopy (XPS) and charge-carrier transport measurements to highlight the improvement in morphological and electrical transport properties in graphene processed with 2SC. While AFM and X-ray photoelectron spectroscopy (XPS) measurements confirm the effectiveness of 2SC in removing PMMA residues, Raman spectroscopy indicates graphene strain and doping reduction. Electrical transport measurements performed on a total of 50 graphene HBs fabricated using 2SC show average hole mobility $\mu_h$ ~7500 m$^2$ V$^{-1}$ s$^{-1}$ and average electron mobility $\mu_e$ ~6300 m$^2$ V$^{-1}$ s$^{-1}$, *i. e.* an improvement of 65% and 37%, respectively, compared to samples processed with 1SC.

## 2. Experiment and methods

Single-crystal graphene arrays[31,32] with a lateral size of 200-250 μm were synthesized via CVD on 2×2 cm$^2$ electropolished Cu-foils (25 μm thick, Alfa Aesar, purity 99.8%) by following the procedure reported by Miseikis et al.[33] Specifically, graphene was synthesized at a temperature of 1060 °C in a cold-wall CVD reactor (Aixtron BM) under argon (Ar), hydrogen (H$_2$) and methane (CH$_4$) with flow ratio of 900:100:1, respectively. Afterwards, the graphene crystals were transferred on highly-doped Si substrates with a 285 nm layer of SiO$_2$ (Siltronix) using a semi-dry technique as reported previously.[33,34] A poly (methyl methacrylate) (PMMA) layer was used to support the graphene single-crystals while detaching them from Cu-foil via electrochemical delamination.[35,36] The PMMA-coated graphene single crystals were then finally deposited on the target SiO$_2$/Si substrate using a micrometric mechanical stage. More details about the graphene-transfer technique can be found in the Supplementary Information (SI). After transferring graphene from Cu to SiO$_2$/Si, a wet chemical cleaning method was used to remove the PMMA layer. For 1SC, the graphene sample was immersed in acetone for 2 hours and rinsed in isopropyl alcohol for 5 minutes, then dried under compressed nitrogen flow. In the case of 2SC, the steps of 1SC were followed by a 3 min bath in remover AR 600-71 and a 10 sec rinse in deionized water, followed by drying with compressed nitrogen. AR 600-71 (Allresist) is a two-component solvent (70%, 1,3-Dioxolane and 30%, 1-Methoxy 2-Propanol), effective at stripping PMMA, Chemical Semi Amplified Resist (CSAR) and novolac-based resists.[37] The two-step cleaning procedure was used after graphene transfer as well as after each fabrication step where PMMA removal was involved, *i. e.* after graphene etching and metal lift-off. During preliminary tests, 1,3-Dioxolane and 1-Methoxy 2-Propanol (Sigma-Aldrich) were also used separately to assess the efficacy of the solvent constituents.

To elucidate the morphology of the graphene surface, atomic force microscopy (AFM) was performed with a Dimension ICON-PT (Bruker). Topographic images were obtained in peak force tapping mode

(Bruker Scan-Asyst).[38] Gwyddion software was used to process the AFM images, extract the surface profile and to perform surface roughness calculations and particle analysis.

Raman spectroscopy was performed with a Renishaw InVia system with a 532 nm laser, and a 100× objective, giving a spot size of ~0.8 µm.[8] Laser power was set to ~1 mW to minimize heating. Raman mapping was performed over an area of 12x12 µm$^2$ with a step size of 1 µm.

XPS analyses of the transferred graphene crystal were conducted on a Thermo Fisher Scientific Escalab 250 xi, equipped with a monochromatic Al-Kα anode (1486.6 eV). Two complementary approaches were used to characterize the samples before and after the 1SC and 2SC process: parallel XPS Imaging for flakes localization and selected Small-area XPS (SAXPS) for spectroscopy. Parallel XPS Imaging is an acquisition mode using a large X-ray spot (i.e. 900 µm). Photoelectrons from the entire defined field of view (250 or 500 µm) are simultaneously collected on the 2D detector. Electrons of a given kinetic energy are focused on the channel-plate detector to produce a direct image of the sample without scanning. By integrating images from consecutive energies, an average spectrum of the considered area can be generated. Maps were recorded in the energy range of C1s, O1s and Si2p, using a 200 eV pass energy and 0.1 eV energy step between each acquisition. SAXPS was performed on the flakes evidenced by the previous map. This method maximizes the detected signal coming from a specific area (60 µm here) and minimizes the signal from the surrounding area. It is achieved by using irises and the spectrometer's transfer lens to flood the area with X-rays but limit the area from which the photoelectrons are collected. High-energy resolution spectral windows of interest were recorded for C1s, O1s and Si2p core levels. The photoelectron detection was performed using a constant analyser energy (CAE) mode (10 eV pass energy) and a 0.05 eV energy step. All the associated binding energies were corrected with respect to the C1s at 284.5 eV.

Electrical transport properties of CVD graphene were investigated by fabricating a total of 78 HBs using electron beam lithography (EBL) on two substrates. 28 HBs were fabricated using 1SC and 50 HBs by using 2SC after each fabrication stage. The HBs were defined using electron-beam lithography (EBL) at 20 kV using a Zeiss UltraPlus scanning electron microscope (SEM) and Raith Elphy Multibeam EBL system. The patterns were defined in positive e-beam resist (PMMA 950k, Allresist GMBH). Graphene was etched using oxygen plasma in a parallel-plate reactive ion etching (RIE) system (Sistec) at 35W with Ar/O$_2$ flow of 5/80 sccm, respectively. The contacts were deposited by thermal evaporation of 60 nm of Au with a 7 nm Ni adhesion layer. Electrical transport characterization was performed in ambient conditions using a custom-made probe station with tungsten tips on micropositioners. Electric field effect was measured using a pair of Keithley 2450 source-measure units, for a 4-terminal resistance measurement and back-gate sweep.

## 3. Results and discussion

### 3.1. AFM characterization

To evaluate the effectiveness of our cleaning procedure, the surface morphology of graphene after 1SC and 2SC were studied via AFM. A 10×10 µm$^2$ area was selected near the edge of a graphene crystal to allow the analysis of polymer residues on graphene and SiO$_2$ surface. The topographical images of the selected area are shown in figure 1a (1SC) and 1b (2SC). The 1S cleaning protocol leaves nanometre-sized particles on graphene.[39] When subjected to the second cleaning step, a remarkable reduction of polymer residues can be observed, revealing a flat and homogeneous graphene surface with only occasional wrinkles. Figure 1c shows two-line profiles extracted from the same area after 1SC (blue curve) and 2SC (red curve). The former is dominated by surface height variation of 0-3 nm, with a number of surface features reaching the height of almost 10 nm. In the case of 2SC, the surface height variation is much less pronounced, with only a few points reaching a height of ~2 nm, corresponding mostly to the surface height variation of the $SiO_2$ substrate, as can be seen outside of the graphene crystal. RMS roughness of the surface was measured on the graphene-coated area to be ~2.8 nm after 1SC and reduced to ~0.6 nm after 2SC, nearly matching the intrinsic roughness of the $Si/SiO_2$ wafer (~0.5 nm, measured before graphene deposition and declared by the supplier).

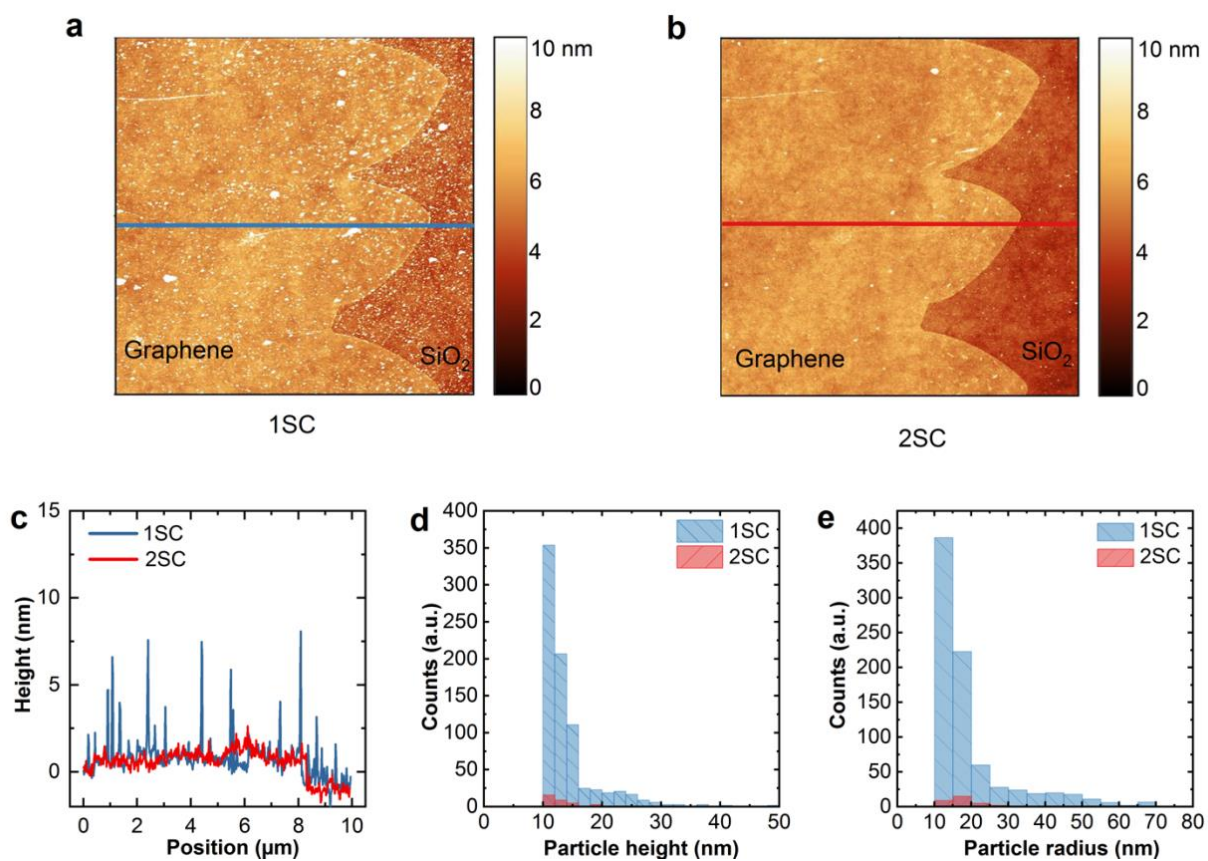

**Fig. 1** (a) Topography image (10×10 µm$^2$) of graphene crystal edge after transfer to $SiO_2/Si$ and 1SC. (b) The same area after the second cleaning step. (c) Surface profile of graphene taken from the topographical images after 1SC (blue) and 2SC (red). (d),(e) Statistical distribution of particle height and radius, respectively, after single- and two-step cleaning.

Particle analysis was done on a 10x10 µm$^2$ area from the centre of a graphene crystal not including bare $SiO_2$ (shown in figure S2a, b in supplementary information). For 1SC, 810 particles were counted, with an average height of ~14±5 nm and an average radius of ~19±13 nm. After the 2$^{nd}$ cleaning step, the number of particles was reduced to 34, with an average height of ~13±3 nm and an average radius of ~20±14 nm.

Statistical distribution of particle height and radius of the sample after each cleaning step is shown in figures 1d and 1e, respectively. These results indicate that >95% of surface contaminants are removed by 2SC compared to standard acetone cleaning. We also note that similar results were obtained on polycrystalline graphene wet transferred onto Si/SiO$_2$ (figure S5). The RMS roughness values obtained from 3×3 µm$^2$ surface of wet-transferred graphene were 2.2 nm after 1SC and 0.7 nm after 2SC. Preliminary experiments were performed to understand whether one of the two constituents of remover AR 600-71, *i.e.*, 1,3-Dioxolane and 1-methoxy-2-propanol, had a major influence on removing particles: we found that the latter is the component yielding cleaner surfaces at AFM analysis, although less effective than remover AR 600-71, indicating that a synergic effect of the two is needed.

### 3.2. Raman analysis

Raman spectroscopy was performed on the sample to estimate graphene quality including doping and strain, after each processing and cleaning step. Figure 2a shows representative Raman spectra obtained after graphene transfer and 1SC (black), 2SC (orange) and full device fabrication (green). This last step corresponds to three rounds of PMMA deposition and 2SC. After 1SC, the spectrum of graphene shows the characteristic Raman G and 2D peaks at ~1586.85 cm$^{-1}$ and ~2680 cm$^{-1}$, respectively, with an average FWHM(2D) of ~27.86 cm$^{-1}$, which can be fitted with a single Lorentzian, as expected for single-layer graphene.[40] The D peak near 1350 cm$^{-1}$ is absent, indicating a negligible amount of defects.[41] To estimate the doping of graphene, we follow the method reported by Basko et al.[42] After transfer, the area ratio of the peaks 2D and G (A(2D)/A(G)) averages ~5.26, corresponding to doping of ~5x10$^{12}$ cm$^{-2}$. The average peak-ratio increases slightly after 2SC, A(2D)/A(G) ~6.0, though a more significant increase is observed after device fabrication, A(2D)/A(G) ~7.9 (figure 2e). This latter value corresponds to doping of ~1x10$^{12}$ cm$^{-2}$, which is consistent with the position of the charge neutrality point (CNP) in field effect measurements, as will be discussed later. We observe a red shift of the average G peak position from 1586.85 cm$^{-1}$ to 1581.9 cm$^{-1}$ which is consistent with a decrease in doping,[43] though we note that Pos(G) is also sensitive to strain.[44] Indeed, we observe a reduction of average 2D width from FWHM(2D) ~27.86 cm$^{-1}$ for graphene after 1SC to FWHM(2D) ~23.6 cm$^{-1}$ for graphene after fabrication, as shown in figure 2f. FWHM(2D) is known to be sensitive to the strain variation within the area of the laser spot[45] and is a good indication of the quality of on-substrate graphene layers.[46] Notably, graphene on SiO$_2$ typically shows higher FWHM(2D) >25 cm$^{-1}$, even for the case of exfoliated flakes.[45,46] This indicates that our ultra-clean graphene presents remarkably low strain fluctuation, which is essential to achieve high carrier mobility. We note that no D peak can be seen after any of the steps, indicating that the process does not induce any measurable amount of defects. Comparison of Raman maps reveals that 2SC is effective at reducing doping in graphene as well as reducing strain inhomogeneity, consistent with the removal of polymeric residues observed in AFM. It should be noted that 2SC is effective not only after graphene transfer but after any of the fabrication steps where PMMA deposition is involved. PMMA deposition and removal with 1SC during any processing step leads to a Raman spectrum resembling that of graphene after transfer (see SI), but treating the sample with 2SC always leads to reduced doping and strain inhomogeneity. Interestingly, repeating the PMMA re-deposition and full 2SC leads to an improvement of Raman characteristics (reduction of FWHM(2D) and red-shift of Pos(G)) compared to as-transferred graphene treated to 2SC, though this effect is not observed after more than 2 re-depositions of polymer. Similarly, treating graphene with AR 600-71 remover beyond the standard 3 minutes does not lead to further improvement. More details regarding Raman and AFM data taken during different processing steps can be found in supplementary information. We also demonstrate that 2SC can be used to remove the PMMA residues from the surface of CVD grown polycrystalline graphene transferred to $SiO_2$/Si using the standard wet etching approach, as shown in supplementary information. Raman (figure S6) and AFM (figure S5) measurements indicate the reduction in doping and removal of PMMA residues, respectively. Finally, also in this case we observed that using separately the two constituents of remover AR 600-71 was less effective than using the commercial product, with 1,3-Dioxolane yielding more sizable improvements in the Raman spectra.

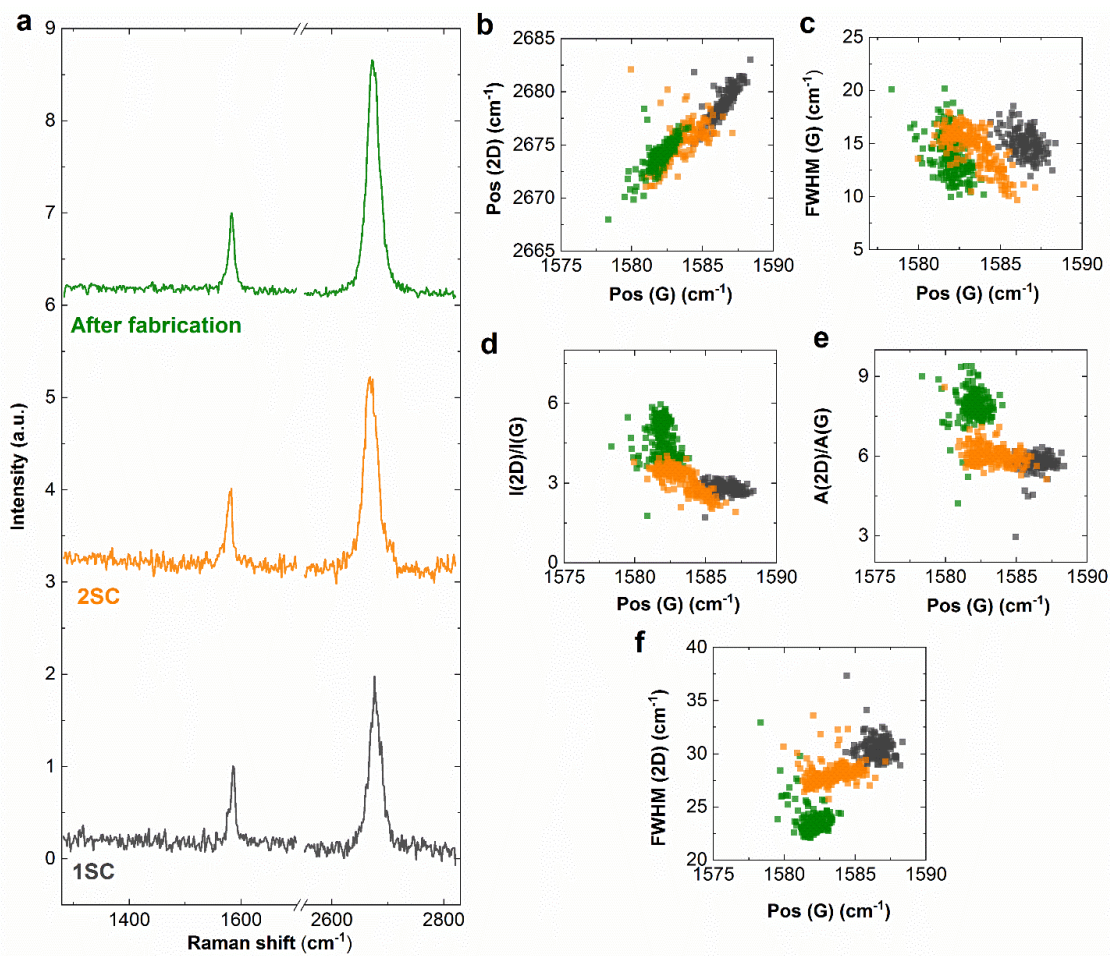

**Fig. 2** (a) Raman spectra of graphene at various stages of processing. (b) Pos (2D) as a function of Pos (G). (c) FWHM (G) as a function of Pos (G). (d) I2D/IG ratio with respect to the position of G-peak. (d) Intensity ratio of 2D and G-peak, (e) Area ratio of 2D and G peaks f) FWHM of 2D-peak as a function of Pos (G).

**3.3 XPS analysis**

Surface chemical composition of graphene after 1SC and 2SC was investigated by XPS. In order to localize the domains, XPS Parallel Imaging (500 x 500 µm²) was used. From the C1s map recorded at 284.5 eV, a 60 µm large area in the middle of a graphene flake was isolated and analysed by selected Small-area XPS (SAXPS). Figure 3 shows the C1s SAXPS spectrum (292-282 eV) recorded on a graphene crystal for 1SC (blue) and 2SC (red) samples. Noting that the red continuous line with a broad asymmetric tail towards higher binding energy mimics a pure $sp^2$-hybridized C-C component (graphene), the C1s spectra suggests that some components at higher energies related to PMMA residues are much more pronounced after 1SC.[39] Indeed, XPS analysis evidences a clear reduction of these spectral components after the second cleaning step, indicating its effectiveness in minimizing PMMA residues. The influence of the remover AR 600-71 is also visible on the C1s XPS maps, which show better-defined flakes after 2SC.

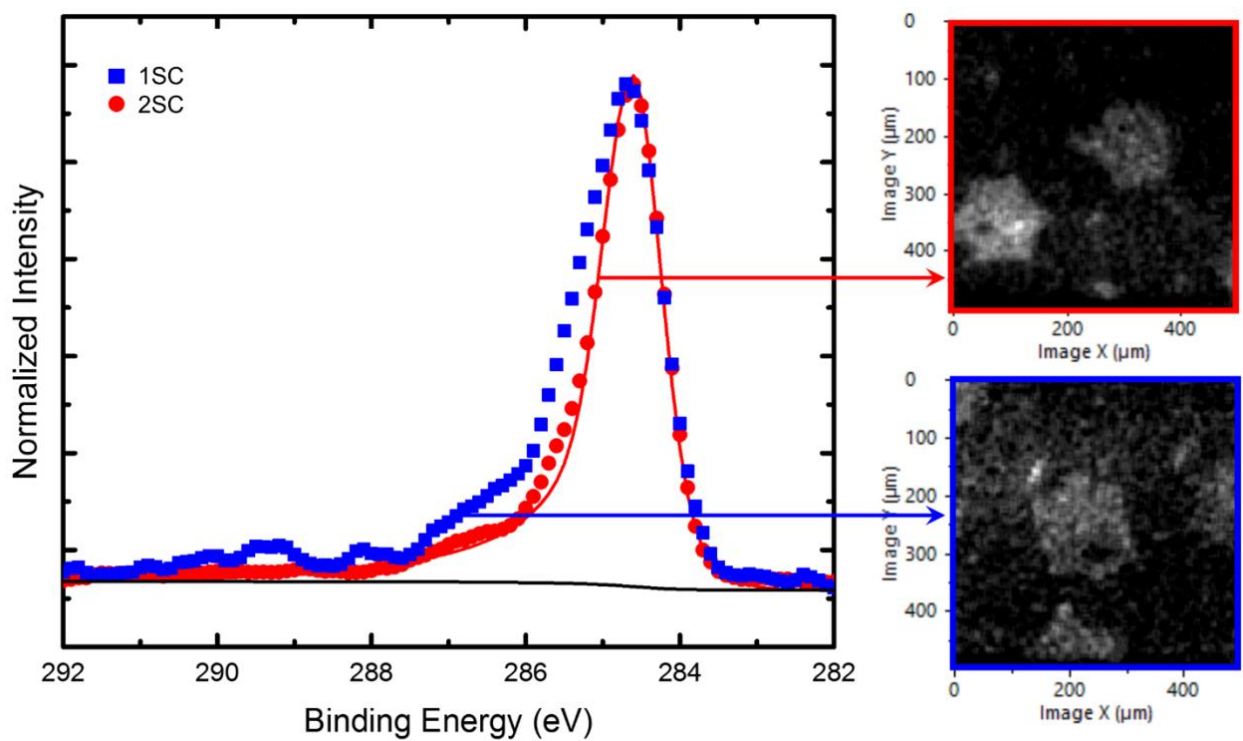

**Fig. 3** C1s SAXPS spectra recorded on 60 µm large area of graphene for samples after 1SC (blue) and 2SC (red). Positions were determined thanks to parallel XPS Imaging (500 x 500 µm²) at 284.5 eV. In the spectra, the red and blue symbols represent the experimental data and the red straight line is related to the fit of a graphene component with an asymmetric shape.

In fact, on the C1s image obtained after 1SC, PMMA residues remain on both the $SiO_2$ substrate and on the graphene crystals, thus yielding a less-contrasted image. In order to ensure that contrasts are mainly related to chemical differences, figure S7 shows C1s and Si2p XPS mapping on the same area. The opposite contrasts observed in the maps indicate that the observed differences are not governed by topographic effects.

## 3.4. Electrical characterization

To investigate the electrical properties of graphene, two sets of back-gated HBs were fabricated using electron beam lithography (EBL). One chip, shown in figure 4a, was processed using 2SC after transfer and each lithography step, whereas the reference chip was fabricated using 1SC at each step.

The field effect response of each HB with the channel aspect ratio L/W=1 was measured as a function of the back-gate voltage using a 4-probe setup, by flowing a current $I_{SD}$ = 1 µA between the longitudinal contacts and measuring the voltage drop $V_{XX}$ along 2 adjacent side contacts. A schematic of the measurement configuration is shown in figure 4b. Resistance was calculated as $R = V_{XX}/I_{SD}$ and the CNP was typically found at values below +15 V. Carrier mobility was calculated for each device as a function of carrier density $n$ (obtained from the applied back-gate bias) using the Drude formula:

$$\mu = 1/(n\,e\,R),$$

where $e$ is the electronic charge. Figure 4c shows the field effect curve obtained for Hall Bar #27 (HB27) on the 2SC chip, and figure 4d shows the resulting carrier mobility as a function of $n$. Hole and electron mobility values at technologically-relevant carrier density of $1\times10^{12}$ cm$^{-2}$ [47] are indicated on the curve. Residual charge density at CNP, $n^*$, was obtained for each device from a linear fit of conductivity on a double-logarithmic scale, as shown in figure 4e for HB27.

Figure 4f shows the measured hole and electron mobility data obtained at $n = 1\times10^{12}$ cm$^{-2}$ for all devices fabricated using 2SC (red and blue dots) and 1SC (black and orange dots). We plot the carrier-mobility values as a function of calculated $n^*$, showing a clear inverse correlation between the charge inhomogeneity and mobility, as observed in high-quality graphene samples.[44] For HBs fabricated using 2SC, average hole mobility $\mu_h$ is estimated to be ~ 7500 cm$^2$ V$^{-1}$ s$^{-1}$, whereas electron mobility $\mu_e$ is ~ 6300 cm$^2$ V$^{-1}$ s$^{-1}$ and average $n^*$ ~ $1.7\times10^{11}$ cm$^{-2}$, with the best values reaching $\mu_h$ ~9100 cm$^2$ V$^{-1}$ s$^{-1}$, $\mu_e$ ~ 7900 cm$^2$ V$^{-1}$ s$^{-1}$ and $n^*$ ~$1.4\times10^{11}$ cm$^{-2}$. The high $\mu$ and low $n^*$ values are consistent with the low FWHM(2D) observed in the Raman measurements. Furthermore, these values compare favourably to the average values obtained from the sample fabricated using 1SC, namely $\mu_h$ ~ 4600 cm$^2$ V$^{-1}$ s$^{-1}$, $\mu_e$ ~ 4500 cm$^2$ V$^{-1}$ s$^{-1}$ and $n^*$ ~ $2.9\times10^{11}$ cm$^{-2}$. The significantly-improved mobility values of 2SC devices indicate that effective removal of polymer residues reduces charge-carrier scattering and unintentional doping.[48,49] This demonstrates that 2SC enables graphene transfer and processing on a large scale with high mobility suitable for the fabrication of functional devices, such as optoelectronic modulators and photodetectors for high-speed telecommunications.[9]

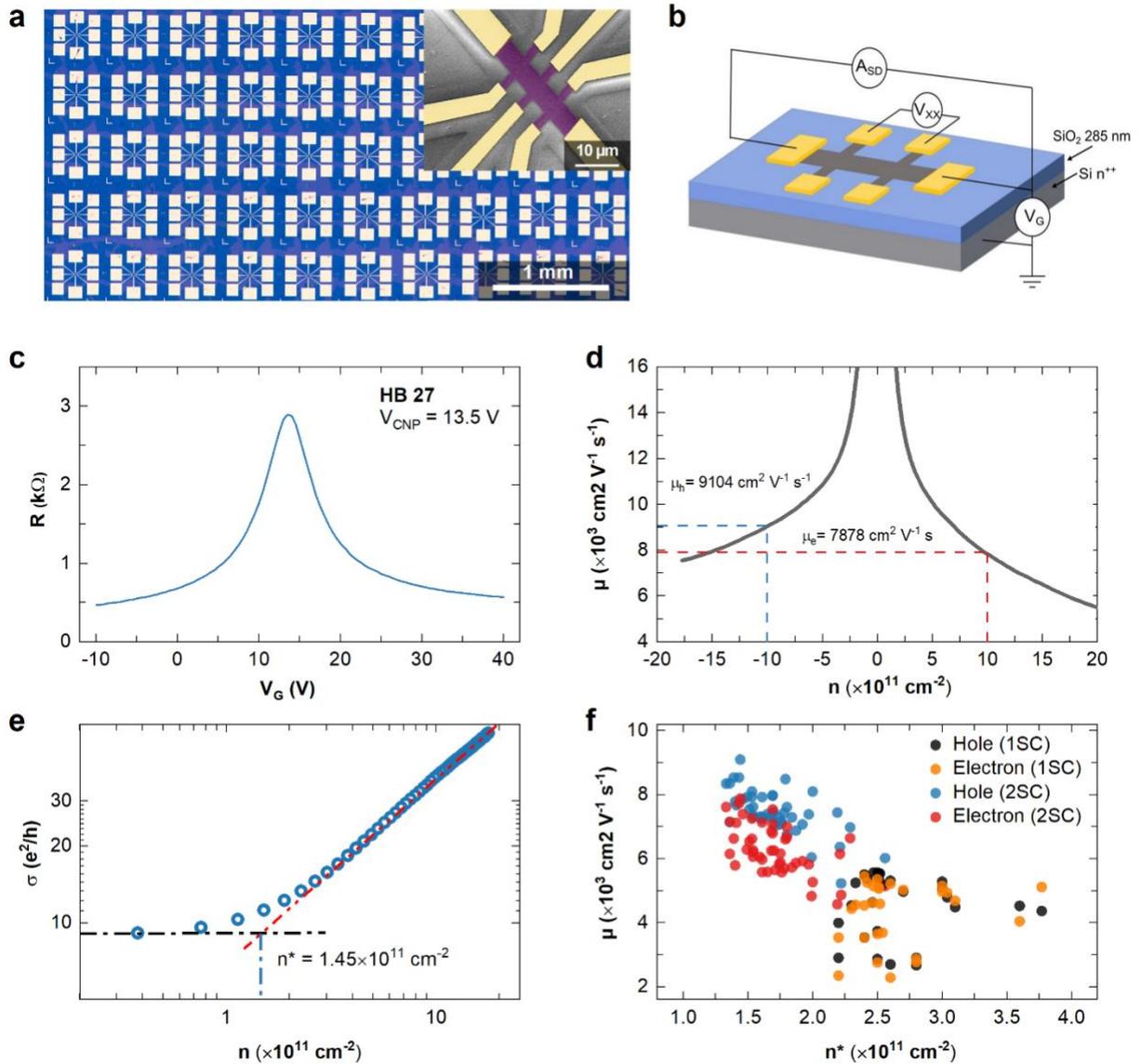

**Fig. 4** Electrical characterization of devices fabricated using 1SC and 2SC. (a) Optical image of 50 graphene Hall bars on SiO$_2$/Si. Inset: false-colour SEM image of a single Hall bar. (b) Schematic diagram of the 4-terminal electrical characterization setup. c) Resistance curve of HB27 prepared with 2SC. (d) Carrier mobility as a function of carrier density calculated from the measurement in (c). e) Linear fit of graphene conductivity as a function of carrier density to estimate the charge inhomogeneity $n^*$. f) Mobility statistics of all graphene Hall bars prepared with 1SC (black, orange) and 2SC (red, blue) as a function of $n^*$.

**Conclusion**

In this work, we demonstrate an effective and rapid two-step cleaning (2SC) method to reduce the polymeric residues present on graphene surface after transfer and lithography processing. In this way, improved mobility CVD graphene can be achieved on SiO$_2$/Si substrates. AFM surface topography measurements and spatially-resolved XPS clearly show the effectiveness of this approach in removing

PMMA residues, for both single-crystal graphene transferred with semi-dry transfer and polycrystalline graphene prepared with wet etching transfer. This makes the approach presented here a relevant technique for preparing high-quality graphene for various applications. A detailed analysis of Raman maps indicates the reduction of graphene doping after 2SC, while the absence of Raman D peak confirms that no structural defects are introduced in graphene. Electrical measurements show a significant improvement of the carrier mobility and residual charge carrier density with respect to chips processed using the traditional fabrication procedure. The 2SC approach does not introduce defects and yields high cleanliness while being scalable, rapid and easy to perform, with a great improvement over the existing approaches such as thermal annealing, scanning probe cleaning, and polymer-free fabrication (stencil mask lithography). Hence, it provides a straightforward route for the achievement of ultra-clean high-mobility scalable graphene devices which are sought after by several applications.

**Conflicts of interest**

There are no conflicts to declare.

**Acknowledgement**

We acknowledge financial support from Fondazione Tronchetti Provera and National Centre for Scientific Research in the framework of the Emergence program at INC-CNRS. The research leading these results has received funding from the European Union Horizon 2020 Programme under Grant Agreement No. 881603 Graphene Core 3.

# Supporting Information

# **Ultra-clean high-mobility graphene on technologically relevant substrates**

Ayush Tyagi[ab], Vaidotas Mišeikis*[be], Leonardo Martini[b], Stiven Forti[b], Neeraj Mishra[be], Zewdu M. Gebeyehu[be], Marco A. Giambra[c], Jihene Zribi[d], Mathieu Frégnaux[d], Damien Aureau[d], Marco Romagnoli[c], Fabio Beltram[a], Camilla Coletti*[be]

[a] NEST, Scuola Normale Superiore, Piazza San Silvestro 12, 56127 Pisa, Italy

[b] Center for Nanotechnology Innovation @NEST, Instituto Italiano di Technologia, Piazza San Silvestro 12, 56127 Pisa, Italy

[c] Photonic Networks and Technologies Lab, CNIT, 56124 Pisa, Italy; CamGraPhIC srl, 56124 Pisa, Italy.

[d] Institut Lavoisier de Versailles UMR 8180 Université Paris-Saclay, UVSQ, CNRS, 78035 Versailles, France

[e] Graphene Labs, Istituto Italiano di Tecnologia, via Morego 30, 16163 Genova, Italy

**Graphene transfer and cleaning process.**

Graphene has been transferred from Cu to Si/SiO$_2$ using the semi-dry transfer method reported previously[S1]. Initially, CVD grown single layer graphene on Cu is coated with a 100 nm PMMA layer and heated at 90°C for 2 minutes. Also, an additional 1.5 μm PPC layer is employed for a stronger mechanical support to the graphene layer[S1]. The graphene/PMMA/PPC stack is then heated again to 90°C for 2 minutes. Furthermore, a PDMS frame is attached to the edge of the Cu foil. SLG electrochemical delamination is then performed in 1 M NaOH. Cu/SLG is used as the anode, and ~2.4 V is applied with respect to a Pt counter electrode. The voltage is controlled to keep the current ~3 mA to avoid excessive formation of H$_2$ bubbles, which may cause damage to SLG. The freestanding polymer/SLG membrane is then removed from the electrolyte, rinsed 2 times in DI water, then dried in air. Using a custom-built aligned lamination setup, the graphene/polymer stack is then transferred on a SiO$_2$/Si wafer heated to 90 °C. The PDMS is then peeled off and the polymer coating is removed by leaving the sample in acetone for 2 hours, then in isopropyl alcohol for 5 minutes, and finally dried under compressed nitrogen flow (1SC). To remove the remaining PMMA residues, the sample was immersed in remover AR600-71 for 3 minutes, rinsed in deionized water for 10 seconds, and finally N$_2$ blow-dried (second part of 2SC).

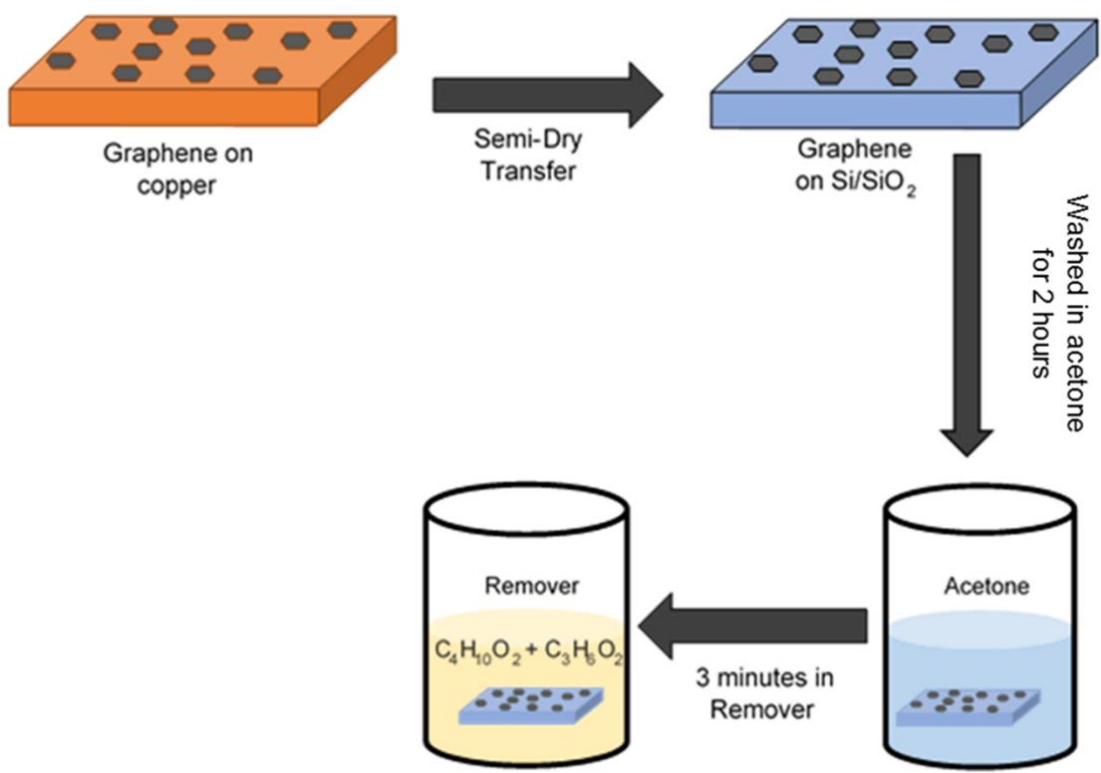

**Fig. S1** Schematics of the transfer and cleaning process of CVD graphene.

**Additional AFM analysis after 1SC and 2SC**

Fig. S2a, b shows AFM topography images obtained at the center of the graphene crystal after 1SC and 2SC, used for the particle analysis presented in Fig. 1d, e of main text.

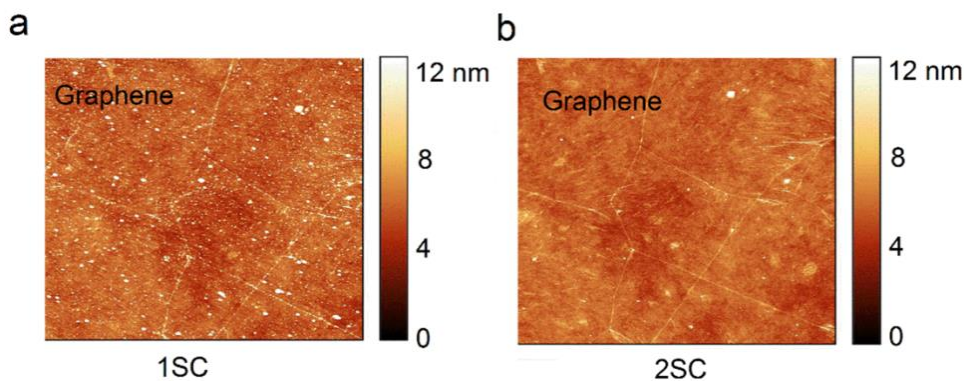

**Fig. S2** (a) AFM topography image (10×10 µm$^2$) obtained at the center of a transferred graphene crystal after 1SC. (b) AFM topography image of the same graphene area after 2SC.

During device fabrication, RIE (O$_2$/Ar plasma) is used to pattern graphene, employing a PMMA etch mask. We have found that RIE process hardens the PMMA, with 1SC leaving even more residues compared to post-transfer cleaning. Fig. S3a shows an AFM topography image of a patterned graphene structure after 1SC. RMS roughness of graphene surface obtained from an area of 4x4 µm$^2$ is 4.4 nm. As can be seen in Fig. S3b, employing 2SC, graphene surface can be cleaned just as well, with RMS roughness of the same area measured at 0.8 nm. Fig. S3c shows a line profile obtained from the same area of patterned graphene after 1SC (blue) and 2SC (red).

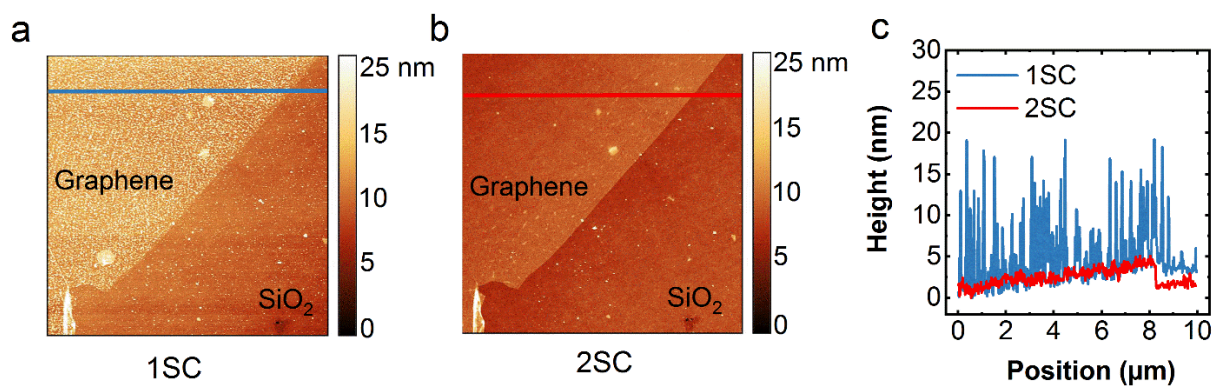

**Fig. S3** (a) AFM topography image (10×10 µm$^2$) of patterned graphene after 1SC. (b) AFM topography image of the same graphene area after 2SC. (c) Line profiles of patterned graphene after 1SC (blue) and 2SC (red).

**Raman analysis for 1S and 2S cleaning during the transfer and lithography process**

For better understanding of the effects of polymer deposition on high quality graphene and subsequent cleaning, we have performed extensive Raman mapping on the same 30x30 µm$^2$ graphene area during device fabrication. A Raman map was obtained after each of the 2 cleaning steps following every graphene fabrication procedure (transfer, patterning and metallization). The collected data is presented in table S1 and Fig. S4. For clarity, in Fig. S4 we present the average values with respective error bars instead of the numerous data points obtained from Raman mapping.

**Table. S1. Average Raman Fit Parameters from Figure S4 (a–e)**

| Graphene cleaning stage | Pos (G) (cm$^{-1}$) | Pos (2D) (cm$^{-1}$) | FWHM (G) (cm$^{-1}$) | FWHM (2D) (cm$^{-1}$) | I(2D)/I(G) | A(2D)/A(G) |
|---|---|---|---|---|---|---|
| Transfer+1SC | 1586.85 | 2680 | 12.13 | 27.86 | 2.27 | 5.26 |
| Transfer+2SC | 1583.35 | 2674.9 | 14.5 | 28.06 | 3.13 | 6 |
| Etching +1SC | 1586.7 | 2678.4 | 9 | 25.7 | 2.3 | 6.7 |
| Etching+2SC | 1582.9 | 2675.2 | 12.2 | 24.6 | 3.6 | 7.3 |
| Lift-off+1SC | 1583.2 | 2673.7 | 11.2 | 25.9 | 3.1 | 7.2 |
| Lift-off+2SC | 1581.9 | 2673.9 | 14.1 | 23.6 | 4.6 | 7.9 |

The general trend of the data is that after each processing step (i. e., transfer, RIE, metallization) and 1SC cleaning, a blue shift of peak positions and a reduction in 2D/G peak intensity and area ratio is observed, compared to pristine graphene. This indicates increasing doping and strain, though this effect is reversed simply by applying the second cleaning step. 2D width, which is correlated to the nanometric strain fluctuations within the laser spot[S3], gradually improves after each step, reaching the lowest value of 23.6 cm$^{-1}$ after metal liftoff and 2SC. This likely indicates that polymer re-deposition releases some of the strain induced in graphene during transfer, without releasing extra contamination due to the effective cleaning by the remover. We note that this 2D width is lower than is generally observed for exfoliated graphene on SiO$_2$.

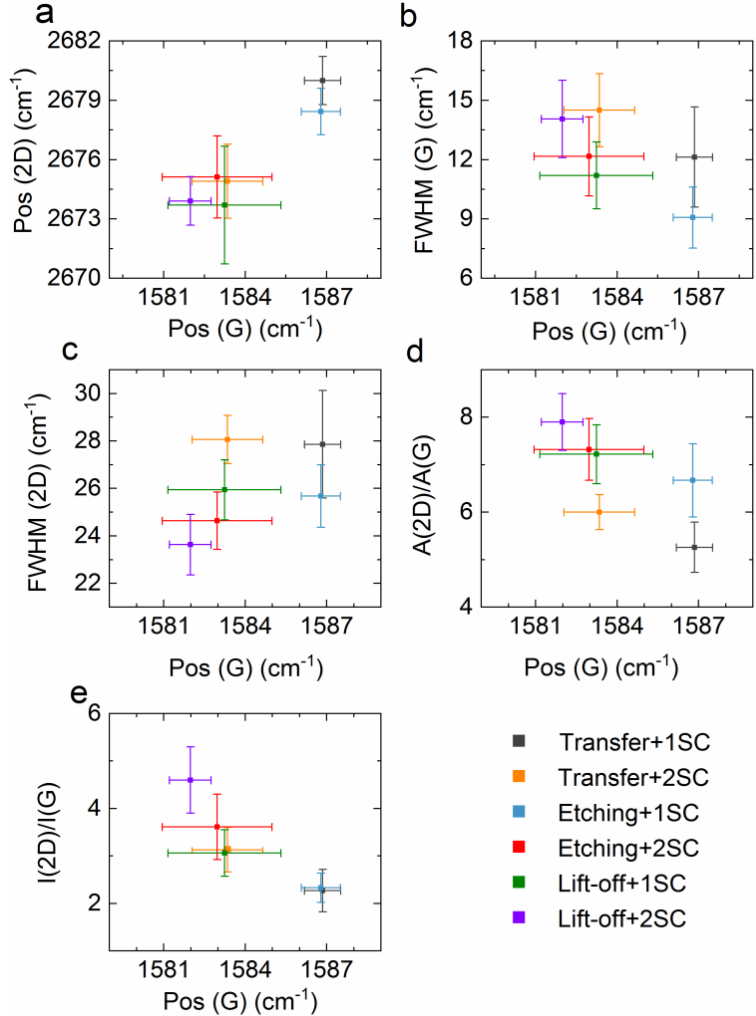

**Fig. S4** Raman correlation plots after 1SC and 2SC at each transfer/processing step. (a) Pos (2D) as a function of Pos (G). (b) FWHM (G) as a function of Pos (G). (c) FWHM of 2D-peak as a function of Pos (G). (d) 2D peak area as a function of Pos (G). (e) 2D peak intensity as a function of Pos (G).

**AFM and Raman analysis of polycrystalline graphene after 1SC and 2SC.**

We demonstrate that graphene processing using 2SC can be used not only for single-crystal graphene after semi-dry transfer, but also for wet-transferred polycrystalline graphene.

Polycrystalline monolayer graphene was grown on copper foil (Alfa Aesar 46365) as reported previously[S4]. It was coated with a 200 nm layer of PMMA (AR-P 679.02, Allresist). Cu foil was etched using a copper etchant solution (30 g/L Ammonium persulfate in $H_2O$, Sigma-Aldrich), leaving a membrane of PMMA/Graphene floating on the surface of the etchant solution. The membrane was rinsed 3 times in deionized water to remove etchant residues. Si/$SiO_2$ was then used to pick up the membrane from the water and dried in ambient conditions for 1h. The sample was then baked on a hotplate at 120°C for 15 minutes to improve graphene adhesion. Subsequently, the PMMA was removed using 1SC, followed by AFM and Raman characterization. The second cleaning step was then performed and the sample was again characterized using AFM and Raman spectroscopy.

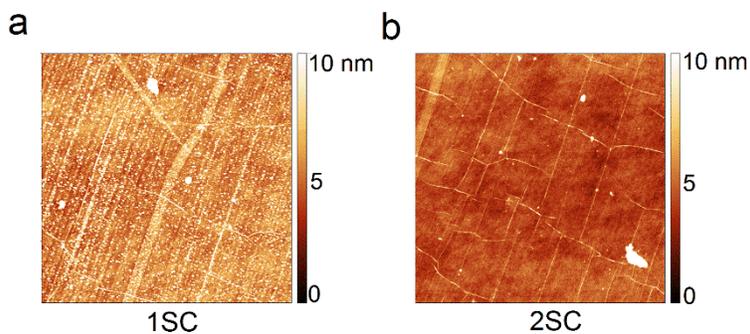

**Fig S5.** (a) AFM topography image (6×6 μm$^2$) taken after transferring polycrystalline graphene on SiO$_2$/Si after 1SC and (b) 2SC.

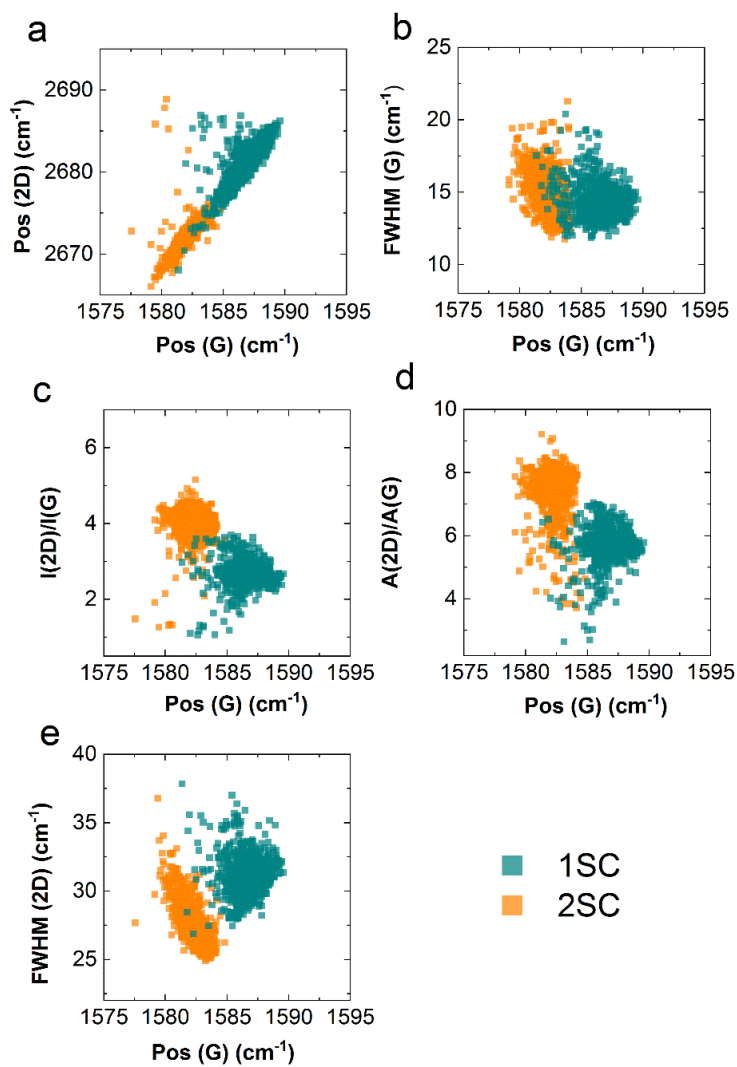

**Fig. S6** (a) Pos (2D) as a function of Pos (G). (b) FWHM (G) as a function of Pos (G). (c) I2D/IG ratio with respect to the Pos (G). (d) Area ratio of 2D and G peaks and e) FWHM of 2D-peak as a function of Pos (G) of polycrystalline graphene after 1SC and 2SC.

As is visible from AFM and Raman analyses, the effect of 2SC in removing polymer residues, reducing strain, and doping, is similar also when applied to continuous polycrystalline graphene, and not only for graphene single crystals.

**XPS parallel imaging for 2S cleaning taken in the C1s and Si2p region**

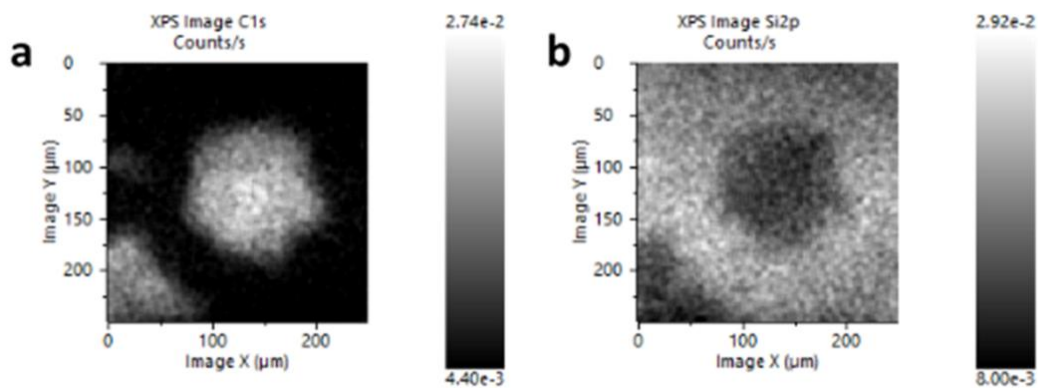

**Fig. S7**  Parallel XPS Imaging of a graphene crystal after transfer to SiO$_2$/Si and 2SC measured at binding energies of (a) 284.5 eV and (b) 102.5 eV.

In Fig. S7, we show two different images taken on the same region (250x250 µm$^2$) of a sample, centered on an individual flake after 2 SC at different binding energies in order to obtain a chemical map of C1s and Si2p intensities. In Fig. S7(a) the higher intensity of carbon gives a clear view of the flake. Nevertheless, such contrast might be related to topographic effects. However, in Fig. S7(b), the intensity of the Si2p peak related to the Si substrate is obviously higher outside of the flake. Such data suggest that the Si2p peak intensity is screened by the graphene flake. This indicates that, indeed, chemical mapping is performed.

**Supporting references**